\documentclass{cimento}

\usepackage{graphicx}  % got figures? uncomment this

\title{Nanosecond pulsed proton beam: upgrade of the accelerator-based neutron facility HiSPANOS at CNA (Spain)}
\shorttitle{Pulsed beam in HiSPANOS neutron facility}
\subtitle{Pulsed beam in HiSPANOS neutron facility}
\author{M.~Mac\'ias\from{ins:x}\from{ins:y}\thanks{Corresponding author. Email address: mmacias4@us.es (M. Mac\'ias).}\ETC,
B.~Fern\'andez\from{ins:x}\from{ins:y},
J.A.~Labrador\from{ins:y},
A.~Romero\from{ins:y}
        \atque
J. Praena\from{ins:z}.} %\thanks{Any footnote to author}
\instlist{\inst{ins:x} Universidad de Sevilla, Seville, Spain
  \inst{ins:y} Centro Nacional de Aceleradores (U. Sevilla, J. Andaluc\'ia, CSIC), Seville, Spain
  \inst{ins:z} Universidad de Granada, Granada, Spain}
\begin{document}

\maketitle

\begin{abstract}
The 3 MV Tandem Pelletron accelerator at the Spanish Accelerator Laboratory (CNA) has been recently upgraded to produce pulsed ion beams for
neutron Time-Of-Flight (TOF) measurements. The upgrade has consisted of two actions: a pulsing system installed at the low energy part of the Tandem accelerator and a new line fully equipped. The pulsing system provides approximately one nanosecond pulse width of protons with variable repetition rates from kHz down to MHz. 
The new line is equipped with conventional devices and a pick-up for timing measurements with high resolution. 
The properties of the whole system have been tested under various working conditions and they are described in some detail.
\end{abstract}

\section{Introduction}

The Centro Nacional de Aceleradores (Spanish Accelerator Laboratory, CNA) was created two decades ago and it is located in Seville~\cite{cna}. The main accelerator is the $3~$MV Tandem model 9SDH-2 of NEC (National Electrostatic Corp.). Few years ago some developments were performed at the Tandem for producing neutron beams in continuous mode given rise to HiSPANOS 
(HiSPAlis NeutrOn Source), the first accelerator-based neutron facility in Spain. Since then, some experiments in continuous wave (CW) have been carried out with different neutron production reactions. Briefly, the $^{7}$Li(p,n)$^{7}$Be reaction has been used for producing different stellar neutron beams for nuclear astrophysics studies
~\cite{nima-ta}\cite{nds-tb}. Also, this reaction has been used for producing thermal neutron beams for dosimetry studies in 
the conventional radiotherapy with photons~\cite{radonco}\cite{ari-leti}. The $^{2}$H(d,n)$^{3}$He reaction has been used for studies of soft errors rate in 
SRAM memories~\cite{micro}. In CW, the experiments are known as integral experiments and the 
measured cross-sections are known as spectrum average cross sections (SACS). SACS are very important quantities in many 
fields and they are considered in the studies of neutron standards evaluations~\cite{nds-18}.
 Nevertheless, if the neutron beam is produced with a pulsed ion beam, time information is added to energy information, thus, neutron velocities may be measured, therefore, neutron-induced cross sections or neutron spectra can be determined as a function of the neutron energy. Since several decades, different pulse systems have been developed for low energy accelerators \cite{moak}. 

\begin{figure}[htbp]
\begin{center}
\includegraphics[scale=0.28]{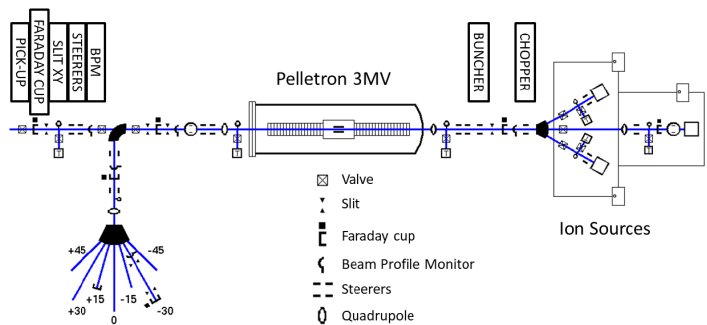} 
\caption{Schematic layout of the $3~$MV Tandem model 9SDH-2 at CNA. The layout includes the ions sources, the chopper, the buncher, 
the tank of $3~$MV, and the different devices of the TOF line: BPM (beam profile monitor), steerer, slit, Faraday cup and pick-up. 
Downstream the $90^{\circ}$ analyzing magnet are located the former lines at $0^{\circ}$, $\pm 15^{\circ}$, $\pm 30^{\circ}$ and 
$\pm 45^{\circ}$. }\label{tandem}
\end{center}
\end{figure}

This paper describes the characteristics and performance of the pulsing 
system for protons, which is able to provide proton bunches of $1~$ns pulse width and repetition rates from $62.5~$kHz to $2~$MHz. 
It satisfies the requirements for measuring at the Tandem accelerator neutron spectra or neutron-induced cross-sections in the 
keV region by means of the TOF technique with a reasonable duration of the experiments.
The neutron beams will be generated with the reaction $^{7}$Li(p,n)$^{7}$Be with protons near its threshold ($1.88~$MeV). 
These requirements are specific for the facility and the considered reaction, thus, they take into account the characteristics of the accelerator in
 terms of maximum proton current available and the yield of the reaction at the mentioned energies. 
Also, the new accelerator line of the Tandem at CNA for TOF measurements is described. Figure~\ref{tandem} shows the location
of both upgrades. 

\section{Pulsing system: chopper and buncher}

The chopping system has one pair of deflection plates installed at the exit of the ion source. When the chopper is on, one plate is always polarized with a $0.5~$kV dc voltage deflecting the beam onto a dump. The other plate is connected to a Fast High Voltage Transistor Switch HTS 31-GSM by Behlke Electronic GMBH. During a period time of one oscillation determined by the Behlke the same $0.5~$kV dc voltage is applied to the other plate, thus, in this period, the beam is allowed to pass the aperture resulting in a pulsed beam. The frequency of the Behlke switch can be varied according to $2^{-n} \times 4~$MHz, $n = 1,2,...,6$. The pulsed beam provided by the chopper has a very long duration, typically tens of nanoseconds, so, it must be compressed in time by the buncher.

The bunching unit is made of a pair of tubular electrodes mounted coaxially to the ion beam. It is setup between the chopping system and the accelerator tank. A radio frequency (RF) voltage of 8 MHz  is supplied to the electrodes. 
Both, the entrance and the exit gap of the tubular bunching electrodes are used for the time compression of the beam pulse. 
At the forefront of the entering pulse, the energy is decreased by the action of the electrodes.
Subsequently, the polarity is inverted, and the energy of protons of the tail of the pulse is increased. Usually,
an entering pulse of tens of nano seconds duration is compressed to $\approx 1~$ns.

\section{Beam line for TOF measurements}

\begin{figure}[htbp]
\begin{center}
\includegraphics[scale=0.12]{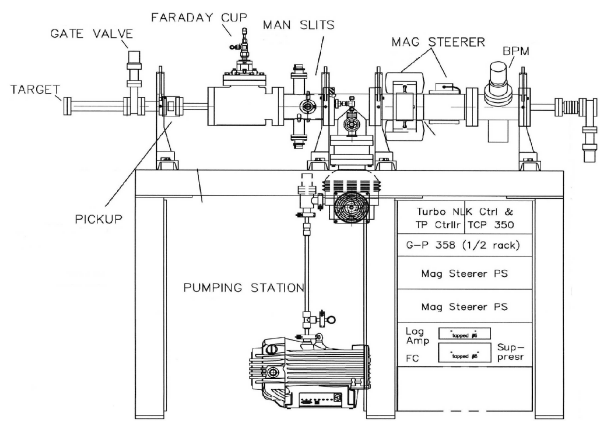} 
\caption{Schematic layout of the TOF line (right to left beam direction).}
\label{line}
\end{center}
\end{figure}

Downstream the acceleration tank the pulsed proton beam enters in the TOF line. The first device is the beam profile monitor (model BMP80). It provides a display of the beam cross sectional shape and position without a significant distortion of the transmission. Then the beam passes through the magnetic steerer, which consists of two units of two coils, each one provides the control in one direction. The third important device is the slit $XY$ (model BDS8), which is suitable to control the beam size in step of $0.01~$mm. Continuing downstream, the beam reaches the Faraday cup (model FC18), which can be used for measuring the beam current. All these devices have been provided by NEC and they can be found in~\cite{nec}. Before a gate valve and neutron production target, the last element is the pickup. It has been designed in collaboration with NTG~\cite{ntg}. The capacitive pickup has been designed as a ring shaped phase probe of $50~\Omega$ impedance. The output signal of the pickup gives information of the integral charge per pulse, the frequency of the pulses, the time width and the noise-to-signal ratio. Figure~\ref{line} shows a schematic layout of the line. 

\section{Operation and performance testing of the pulsing system}

\begin{figure}[htbp]
\begin{center}
\includegraphics[scale=0.25]{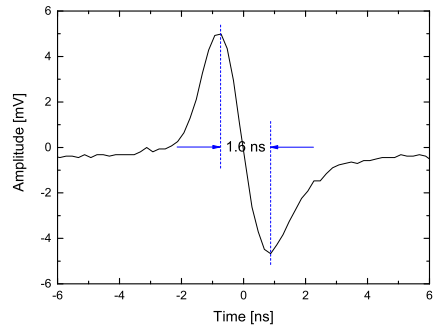} 
\caption{The signal induced in the pickup when a proton pulse crosses it shows a noise-to-signal ratio of $0.5~\%$, the proton beam featuring an average current of $60~$nA and a mean frequency of $62.5~$kHz.}
\label{pu}
\end{center}
\end{figure}

The goal of the tests is to determine the pulse width, beam current and pulse frequency. Initially, the continuous proton beam was transmitted until the TOF line. The 
measured current was $30~\mu$A. Then, the beam was pulsed with a repetition rate of $62.5~$kHz and the buncher was also turn on. 
In the best configuration the measured current at the TOF line was $60~$nA and the induced signal in the pickup was detected, 
see Figure~\ref{pu}. The noise-to-signal ratio was $0.5~\%$, a mean frequency of $62.5~$kHz was measured 
and a time width peak to peak was $1.6~$ns.

\section{Conclusions}

The pulsing system installed in the $3~$MV Van de Graaff accelerator has improved pulsed beams for neutron 
time-of-flight measurements. In the optimal configuration is possible to deliver pulsed proton beams of 1.6 ns pulse width with a repetition
rate of 62.5 kHz, which provides at the end of the TOF line an average current of 60 nA. 
The outlook of this works consists of further optimization of synchronization between chopper and buncher and the possible reduction of 
the pulse width. 

\acknowledgments
This work was supported by the Spanish projects FPA2013-47327-C2-1-R, FPA2014-53290-C2-2-P, FPA2016-77689-C2-1-R, J. de  Andaluc\'{\i}a P11-FQM-8229, FIS2015-69941-C2-1-P (MINECO-FEDER, EU), AECC-PS16163811PORR and the funding agencies of the participating institutes.


\begin{thebibliography}{0}

\bibitem{cna} Garc{\'i}a L{\'o}pez J. $et$ $al.$, $Nucl.$ $Instrum.$ $Meth.$ $Phys.$ $Res.$ $B$, \textbf{161-163} (2000) 1137-1142.

\bibitem{nima-ta} Praena J. $et$ $al.$, $Nucl.$ $Instrum.$ $Meth.$ $Phys.$ $Res.$ $A$, \textbf{727} (2013) 1-6.

\bibitem{nds-tb} Praena J. $et$ $al.$, $Nucl.$ $Data$ $Sheets$, \textbf{120} (2014) 205-207.

\bibitem{radonco} Praena J. $et$ $al.$, $Radiother.$ $Oncol.$, \textbf{115-S375} (2015).

\bibitem{ari-leti} Irazola L. $et$ $al.$, $Appl.$ $Radiat.$ $Isotopes$, \textbf{107} (2016) 330–334.

\bibitem{micro} Malag{\'o}n D. $et$ $al.$, $Microelectron.$ $Reliab$, \textbf{78} (2017) 38-45.

\bibitem{nds-18} Carlson A. D. $et$ $al.$, $Nucl.$ $Data$ $Sheets$, \textbf{148} (2018) 143–188.

\bibitem{moak} Moak C. D. $et$ $al.$, $Rev.$ $Sci.$ $Instrum.$, \textbf{35} (1964) 672-679. 

\bibitem{nec} $http://www.pelletron.com/$

\bibitem{ntg} $https://www.ntg.de/$



\end{thebibliography}
\end{document}